\documentclass[conference]{IEEEtran}
\usepackage{amsmath,amsopn,amssymb,bm}
\usepackage{amsfonts}
\usepackage{gensymb}
\usepackage{amsthm} 
\usepackage{graphicx,xspace,color}
\usepackage{epsfig}
\usepackage{longtable,multirow}
\usepackage{array}
\usepackage{stfloats}
\usepackage{cuted}
\usepackage{cite}
\usepackage[nolist]{acronym}
\usepackage{soul}
\usepackage{algorithm,algpseudocode}
\usepackage{algcompatible}
\usepackage{enumerate}
\usepackage{lettrine}
\usepackage{mathtools}
\usepackage{comment}
\usepackage{subcaption}

\usepackage[font=small]{caption}

\usepackage{booktabs}
\graphicspath{ {./figures/}}

\newtheorem{remark}{Remark}

\newtheorem{lemma}{Lemma}
\newtheorem{proposition}{Proposition}
\newtheorem{example}{Example}

\begin{acronym}
	\acro{THz}{terahertz}
	\acro{UPA}{uniform planar array}
	\acro{SW}{spherical wave}
	\acro{MIMO}{multiple-input multiple-output}
	\acro{FP}{favorable propagation}
	\acro{IRS}{Intelligent reflecting surface}
	\acro{IRSs}{intelligent reflecting surfaces}
	\acro{EM}{electromagnetic}
	\acro{MIMO}{multiple-input multiple-output}
	\acro{PEC}{perfect electric conductor}
	\acro{Tx}{transmitter}
	\acro{Rx}{receiver}
	\acro{RF}{radio-frequency}
	\acro{LoS}{line-of-sight}
	\acro{SNR}{signal-to-noise ratio}
	\acro{SEP}{surface equivalent principle}
\end{acronym}	

\IEEEoverridecommandlockouts 
\begin{document}
	
\title{Electromagnetic Modeling of Holographic Intelligent Reflecting Surfaces at Terahertz Bands}
\author{\IEEEauthorblockN{Konstantinos Dovelos\IEEEauthorrefmark{1}, Stylianos D. Assimonis\IEEEauthorrefmark{2}, Hien Quoc Ngo\IEEEauthorrefmark{2}, Boris Bellalta\IEEEauthorrefmark{1}, and Michail Matthaiou\IEEEauthorrefmark{2}}
	\IEEEauthorblockA{$^*$Department of Information and Communication Technologies, Universitat Pompeu Fabra (UPF), Barcelona, Spain}
	\IEEEauthorblockA{$^\dagger$Institute of Electronics, Communications and Information Technology (ECIT), Queen's University Belfast, Belfast, U.K.}
	Email: \{konstantinos.dovelos, boris.bellalta\}@upf.edu, \{s.assimonis, hien.ngo, m.matthaiou\}@qub.ac.uk
}

\maketitle

\begin{abstract}
\ac{IRS}-assisted wireless communication is widely deemed a key technology for 6G~systems. The main challenge in deploying an IRS-aided \acf{THz} link, though, is the severe propagation losses at high frequency bands. Hence, a \ac{THz} IRS is expected to consist of a massive number of reflecting elements to compensate for those losses. However, as the IRS size grows, the conventional far-field assumption starts becoming invalid and the spherical wavefront of the radiated waves must be taken into account. In this work, we focus on the near-field and analytically determine the IRS response in the Fresnel zone by leveraging electromagnetic theory. Specifically, we derive a novel expression for the path loss and beampattern of a holographic IRS, which is then used to model its discrete counterpart. Our analysis sheds light on the modeling aspects and beamfocusing capabilities of \ac{THz} \ac{IRS}s.
\end{abstract}

\begin{IEEEkeywords}
Beamfocusing, electromagnetics, intelligent reflecting surfaces, near-field, THz communications. 
\end{IEEEkeywords}

\section{Introduction}
To overcome the imminent spectrum scarcity, \acf{THz} communication is favored for 6G wireless networks because of the abundant spectrum available in the \ac{THz} band (0.1 to 10 THz)~\cite{6G_networks}. However, \ac{THz} links suffer from high propagation losses, and thus transceivers with a massive number of antennas are needed to compensate for those losses by means of sharp beamforming~\cite{5G_prospective}. On the other hand, the power consumption of \ac{THz} \ac{RF} circuits is much higher than their sub-6 GHz counterparts, which might undermine the deployment of large-scale antenna arrays in an energy efficient manner~\cite{THz_powerconsumption}. Consequently, addressing these engineering challenges is of paramount importance for future \ac{THz} communication systems. 

Looking beyond conventional antenna arrays, the advent of metasurfaces, which can customize the behavior (e.g., reflection, absorption, polarization, etc.) of \ac{EM} waves, has paved the way for novel wireless technologies, such as \ac{IRSs}~\cite{smart_radio_environment}. Specifically, an \ac{IRS} consists of nearly passive reconfigurable elements that can alter the phase of the impinging waves to reflect them toward a desired direction~\cite{relay_vs_irs}.

There is a large body of literature that investigates the modeling and performance of \ac{IRS}-aided systems at the sub-6~GHz and millimeter wave bands. Nevertheless, the majority of those works, e.g., \cite{irs_mmwave_placement,thz_irs_coverage, em_based_model1,em_based_model2} and references therein, focus on the far-field regime, where the spherical wavefront of the emitted \ac{EM} waves degenerates into a plane wavefront. Although the far-field assumption facilitates mathematical analysis, it might not be valid for \ac{IRSs} operating at the \ac{THz} band. In particular, an electrically large~\ac{IRS} must be placed close to the \ac{Tx} or \ac{Rx} in order to effectively compensate for the path loss of the  \ac{Tx}-IRS-\ac{Rx} link. As a result, one of the link ends is likely to operate in the radiating near-field of the \ac{IRS}. Additionally, packing an unprecedented number of sub-wavelength reflecting elements into an aperture yields a so-called \textit{holographic reflecting surface}~\cite{holographic_mimo}, which can offer ultra-narrow pencil beams and extremely large power gains.\footnote{In this paper, \textit{holographic \ac{IRS}} refers to a continuous (or quasi-continuous) passive aperture, akin to~\cite{holographic_mimo}.} A few recent papers~\cite{nf_pathloss_model1,nf_pathloss_model2} proposed a path loss model that is applicable to near-field using the popular ``cos$^{q}$" radiation pattern for each IRS element, but considering a discrete \ac{IRS}. In a similar spirit,~\cite{icc_2021, power_scaling_law_irs} analyzed the power scaling laws and near-field behavior of discrete IRSs modeled as planar antenna arrays; note that \cite{power_scaling_law_irs} derived an upper bound on the near-field channel gain, and hence its applicability is limited. From the relevant work, we distinguish~\cite{thz_holographic_irs}, where the authors showed that the far-field beampattern of a holographic \ac{IRS} can be well approximated by that of an ultra-dense discrete \ac{IRS}. 

To the best of our knowledge, holographic \ac{IRSs} have not yet been studied in the near-field region and for arbitrary \ac{Tx}/\ac{Rx} locations. This paper aims to fill this gap in the literature and shed light on the fundamentals of \ac{THz} \ac{IRSs}. Specifically:
\begin{itemize}
\item We determine the field scattered by a holographic \ac{IRS} in the radiating near-field, i.e., Fresnel zone. More particularly, we employ physical optics from \ac{EM} theory to model the \ac{IRS} as a large conducting plate, and then derive the scattered field in closed-form by exploiting the small physical size of \ac{THz} \ac{IRSs}. 
\item We show that the near-field behavior differs significantly from its far-field counterpart, and hence the derived channel model should be adopted for electrically large \ac{IRSs}. Moreover, the near-field beampattern of a contiguous \ac{IRS} can be accurately approximated by that of an ultra-dense discrete \ac{IRS}, thereby enabling the practical realization of holographic reflecting surfaces.
\item  We discuss the implications of the \ac{EM}-based model and highlight the importance of beamfocusing in single-user and multi-user transmissions. 
\end{itemize}

\textit{Notation}: $\mathcal{A}$ is a set, $\mathbf{A}$ is a vector field, $\mathbf{a}$ is a vector, $\mathbf{e}_x$, $\mathbf{e}_y$, and $\mathbf{e}_z$ denote the unit vectors along the $x$, $y$, and $z$ axes, respectively; $\mathbf{e}_{r}$, $\mathbf{e}_{\theta}$, and $\mathbf{e}_{\phi}$ denote the unit vectors along the radial, polar, and azimuth directions, respectively; $\text{erf}(x) = \frac{2}{\sqrt{\pi}} \int_{0}^xe^{-t^2}dt$ is the error function; $\text{sinc}(x)=\frac{\sin(x)}{x}$ is the sinc function; and $x\sim\mathcal{CN}(\mu,\sigma^2)$ is a complex Gaussian variable with mean $\mu$ and variance $\sigma^2$.

\begin{figure}[t]
	\centering
	\includegraphics[width=0.8\linewidth]{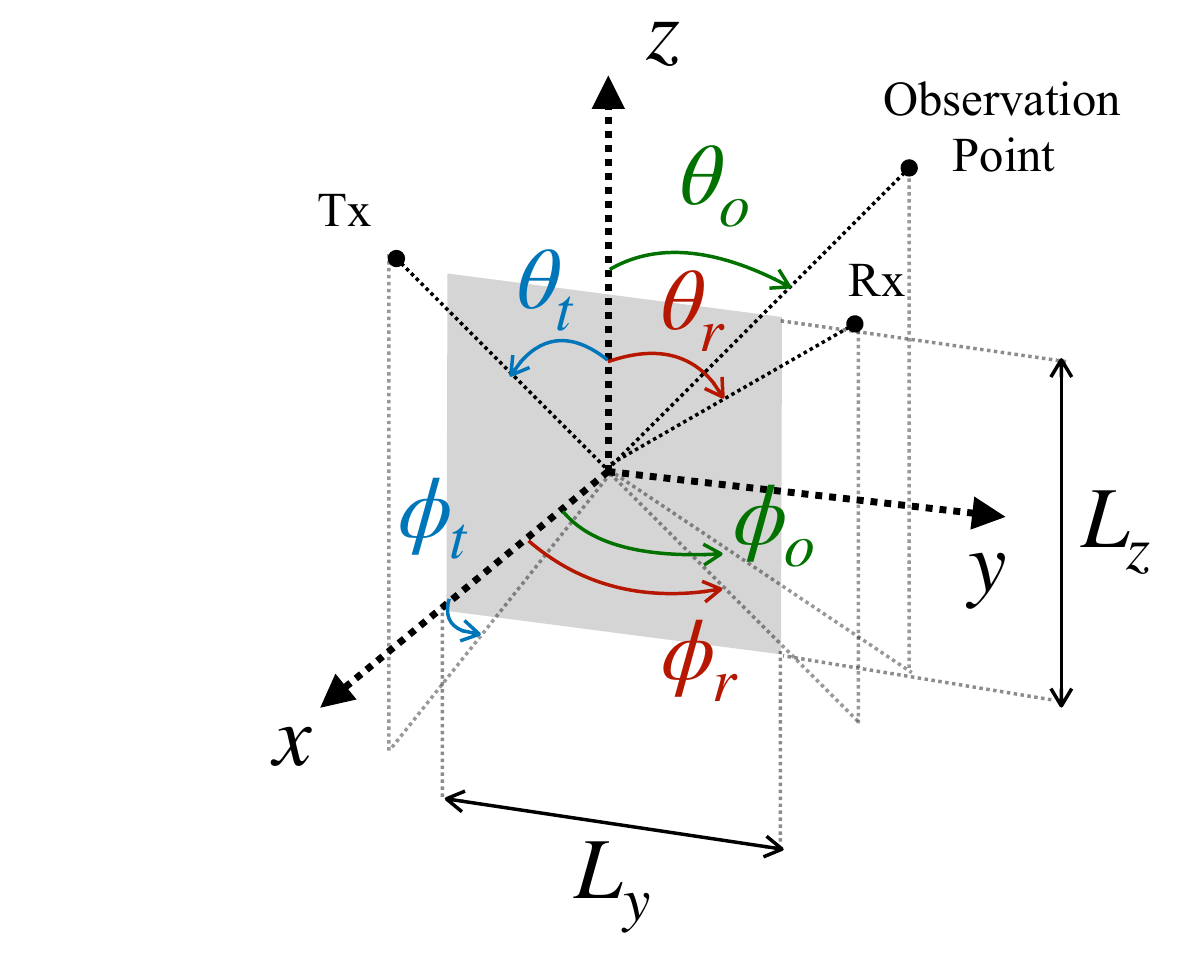}
	\caption{Illustration of the \ac{IRS} geometry under consideration.}
	\label{fig:irs_illustration}
\end{figure}
\section{Electromagnetics-Based Channel Model}\label{sec:channel_model}
Consider a holographic \ac{IRS} of size $L_y\times L_z$, where $L_y$ and $L_z$ denote the dimensions along the $y$ and $z$ directions, respectively. The coordinate system is placed at the center of the \ac{IRS}, as shown in Fig.~\ref{fig:irs_illustration}. Thus, the \ac{IRS} is represented by the planar surface $\mathcal{S}=\{(y,z): |y| \leq L_y/2, |z| \leq L_z/2\}$. In the sequel, we focus on the Fresnel zone of the \ac{IRS}, which refers to all distances $r$ satisfying~\cite{balanis_antenna_theory}
\begin{equation}
0.62 \sqrt{L_{\max}^3/\lambda} < r \leq 2 L^2_{\max}/\lambda,
\end{equation}
where $L_{\max} = \max(L_y,L_z)$ denotes the maximum dimension of the  \ac{IRS}, and $\lambda$ is the carrier wavelength. 

\subsection{Spherical Wavefront}
Consider an infinitesimal dipole antenna emitting a spherical wave; the dipole is placed parallel to the \ac{IRS}. The exact position of the transmit antenna is described by the tuple $(x_t,y_t,z_t) = ( r_t\cos\phi_t\sin\theta_t,r_t\sin\phi_t\sin\theta_t,r_t\cos\theta_t)$, where $r_t$ is the radial distance, whilst $\phi_t$ and $\theta_t$ are the azimuth and polar angles of arrival, respectively. The electric field (E-field) of the spherical wave impinging on the $(y,z)$th point of the \ac{IRS} can be expressed as~\cite{balanis_antenna_theory}
\begin{equation}\label{eq:spherical_wave}
\mathbf{E}_i = E_{\theta}\mathbf{e}_{\theta} = j \sqrt{\frac{\eta P_tG_t}{4\pi}}\frac{e^{-jkr_t(y,z)}}{r_t(y,z)}\mathbf{e}_{\theta},
\end{equation}
where $\eta$ is the wave impedance, $k = 2\pi/\lambda$ is the wavenumber, $P_t$ is the transmit power, $G_t$ is the gain of the transmit antenna,~and 
\begin{align}\label{eq:tx_distance}
&r_t(y,z) \triangleq \sqrt{x_t^2 + (y_t-y)^2 + (z_t-z)^2} \nonumber\\
&= r_t\sqrt{1 + \frac{y^2}{r_t^2} - \frac{2\sin\phi_t\sin\theta_ty}{r_t} + \frac{z^2}{r_t^2} - \frac{2\cos\theta_tz}{r_t}}
\end{align}
is the respective distance. Note that~\eqref{eq:spherical_wave} holds for all distances $r_{t}(y,z) \gg \lambda$, where the radial and azimuthal components $E_r$ and $E_\phi$ of the E-field are approximately zero. From Maxwell's equations, the magnetic field is specified as
\begin{align}\label{eq:H_field_spherical}
\mathbf{H}_i = \frac{j}{\eta k} \nabla\times &\mathbf{E}_i =\frac{j}{\eta k}\frac{1}{r}\frac{\partial (rE_{\theta})}{\partial r} \mathbf{e}_{\phi} \nonumber\\
&= \frac{j}{\eta}\sqrt{\frac{\eta P_tG_t}{4\pi}}\frac{e^{-jkr_t(y,z)}}{r_t(y,z)}\mathbf{e}_{\phi} = H_{\phi}\mathbf{e}_{\phi},
\end{align}
where $r = r_t(y,z)$ in the partial derivative for notational convenience. Owing to the small physical size of \ac{THz} \ac{IRS}s, the amplitude variation $1/r_t(y,z)$ across $\mathcal{S}$ is marginal~\cite{icc_2021}; for example, an electrically large \ac{IRS} of size $200 \lambda \times 200 \lambda$ occupies only $20\times 20$ cm$^2$ at $f=300$ GHz. On the contrary, the phase variation $kr_t(y,z)$ is significant and cannot be ignored. In light of these observations, we henceforth consider
\begin{equation}\label{eq:fresnel_aprox}
\frac{e^{-jk r_t(y,z)}}{r_t(y,z)} \approx \frac{e^{-jk (r_t+\tilde{r}_t(y,z))}}{r_t},
\end{equation}
where $r_t(y,z)\approx r_t + \tilde{r}_t(y,z)$, with 
\begin{align}
\tilde{r}_t(y,z) =  & \frac{y^2(1-\sin^2\phi_t\sin^2\theta_t)}{2r_t} - y\sin\phi_t\sin\theta_t \nonumber\\
&+  \frac{z^2\sin^2\theta_t}{2r_t} - z\cos\theta_t,
\end{align}
which follows from the second-order Taylor approximation $(1+x)^\alpha \approx 1 + \alpha x + \frac{1}{2}\alpha(\alpha-1)x^2$ of \eqref{eq:tx_distance}.

\subsection{Scattered Field in the Fresnel Zone}
According to the surface equivalence principle, the obstacle-free equivalent problem involves an electric current density $\mathbf{J}(y,z)$ (measured in A/m$^{2}$) and a magnetic current density $\mathbf{M}(y,z)$ (measured in V/m$^2$) on $\mathcal{S}$, which satisfy the boundary conditions~\cite[Ch.~7]{balanis_electromagnetics}
\begin{align}
\hat{\mathbf{n}}\times \mathbf{H}|_{x=0} &= \mathbf{J}(y,z), \\
\hat{\mathbf{n}}\times \mathbf{E}|_{x=0} &= \mathbf{M}(y,z) = 0,
\end{align}
where $\mathbf{E} = \mathbf{E}_i + \mathbf{E}_s$ and $\mathbf{H} = \mathbf{H}_i + \mathbf{H}_s$ are the total electric and magnetic fields, respectively, $\mathbf{E}_s$ and $ \mathbf{H}_s$ are the corresponding scattered fields, and $\hat{\mathbf{n}} = \mathbf{e}_x$ is the normal vector of $\mathcal{S}$.\footnote{The E-field inside $S$ is assumed to be zero, akin to the \ac{PEC} paradigm. The PEC model is used for simplicity. Our analysis can readily be applied to the impedance surface model~\cite{modern_sc_theory}.}  Assuming that $\mathcal{S}$ is an infinite \ac{PEC}, it can be replaced by a virtual source with $\hat{\mathbf{n}}\times \mathbf{H}_s = \hat{\mathbf{n}}\times\mathbf{H}_i$, hence yielding $\mathbf{J}(y,z) = 2 \hat{\mathbf{n}}\times \mathbf{H}_i|_{x=0}$.\footnote{We assume that image theory holds for a finite plate. Such an assumption can be made in our case because the dimensions of the \ac{IRS} are very large compared to the wavelength.} Note that the actual \ac{IRS} exhibits a surface impedance, which can change the phase of the surface current density $\mathbf{J}(y,z)$. Thus, we model that property as $\mathbf{J}(y,z) = (2\hat{\mathbf{n}}\times \mathbf{H}_i|_{x=0})e^{j\varphi(y,z)} $~\cite{em_based_model1},\cite{em_based_model2}. The phase shift profile $\varphi(y,z)$ is nonlinear due to the spherical wavefront of the incident wave. To this end, it is decomposed~as
\begin{equation}\label{eq:phase_profile}
\varphi(y,z) = k\left(C_1 y^2 +  C_2 y +  C_3 z^2  + C_4 z\right),
\end{equation}
where $C_1$, $C_2$, $C_3$, and $C_4$ are properly selected constants. 

Let $(x_r,y_r,z_r) = ( r_r\cos\phi_r\sin\theta_r,r_r\sin\phi_r\sin\theta_r,r_r\cos\theta_r)$ be the receiver location, where $r_r$ is the radial distance, while $\phi_t$ and $\theta_t$ denote the azimuth and polar angles of departure, respectively. Next, the scattered E-field at the receiver is analytically determined using the auxiliary vector potential
\begin{align}\label{eq:vector_potential}
\mathbf{A}(x_r,y_r,&z_r) \triangleq \frac{\mu}{4\pi}\int \!\!\!\int_\mathcal{S} \mathbf{J}(y,z)\frac{e^{-jkr_r(y,z)}}{r_r(y,z)} dydz \nonumber \\[0.15cm]
&\overset{(a)}\approx \frac{\mu e^{-jkr_r}}{4\pi r_r}  \int \!\!\! \int_\mathcal{S} \mathbf{J}(y,z)e^{-jk\tilde{r}_r(y,z)} dydz  \nonumber \\[0.15cm]
& = \frac{\mu e^{-jkr_r}}{4\pi r_r} (\tilde{A}_r\mathbf{e}_{r}  + \tilde{A}_\theta\mathbf{e}_{\theta} + \tilde{A}_\phi\mathbf{e}_{\phi }),
\end{align}
where $\mu$ is the magnetic permeability of the propagation medium,~$(a)$ follows from the Fresnel approximation of the distance $r_r(y,z)\approx r_r + \tilde{r}_r(y,z)$, and
\begin{align}
\tilde{A}_r \! &= \! \int\!\!\!\!\int_\mathcal{S} (J_y\sin\theta_r \sin\phi_r + J_z\cos\theta_r )e^{-jk\tilde{r}_r(y,z)} dydz,\\
\tilde{A}_\theta \! &= \!\int\!\!\!\!\int_\mathcal{S} (J_y\cos\theta_r \sin\phi_r - J_z\sin\theta_r)e^{-jk\tilde{r}_r(y,z)} dydz, \label{eq:Atheta} \\
\tilde{A}_\phi \!&= \!\int\!\!\!\!\int_\mathcal{S} J_y\cos\phi_r e^{-jk\tilde{r}_r(y,z)}dydz.\label{eq:Aphi}
\end{align}
Using the radiation equations for any receive distance $r_r \gg \lambda$, we finally have~\cite[Eq. (6.122)]{balanis_electromagnetics} 
\begin{align}
\mathbf{E}_s = -\eta \frac{j k e^{-j kr_r}}{4\pi r_r} (\tilde{A}_{\theta}\mathbf{e}_{\theta} + \tilde{A}_{\phi}\mathbf{e}_{\phi}).
\end{align}
\begin{proposition}
The scattered E-field at the receive position $(r_r\cos\phi_r\sin\theta_r,r_r\sin\phi_r\sin\theta_r,r_r\cos\theta_r)$, when the \ac{IRS} is illuminated by a spherical wave originated from $( r_t\cos\phi_t\sin\theta_t,r_t\sin\phi_t\sin\theta_t,r_t\cos\theta_t)$, is given by 
\begin{align}\label{eq:scattered_efield}
\mathbf{E}_s &= -\frac{L_yL_z}{\lambda}\frac{|E_i| e^{-j k(r_t + r_r)}}{r_r}\cos\phi_t\sin\theta_r S_{yz} \mathbf{e}_{\theta},
\end{align}
where $|E_i| = \sqrt{\frac{\eta P_tG_t}{4\pi r^2_t}}$ is the magnitude of the incident field, and $S_{yz}\in[0,1]$ is the normalized space factor of the \ac{IRS} specified by~\eqref{eq:space_factor} at the bottom of the next page for 
\begin{figure*}[t]
	\centering
	\begin{subfigure}{.5\textwidth}
		\centering
		\includegraphics[width=0.83\linewidth]{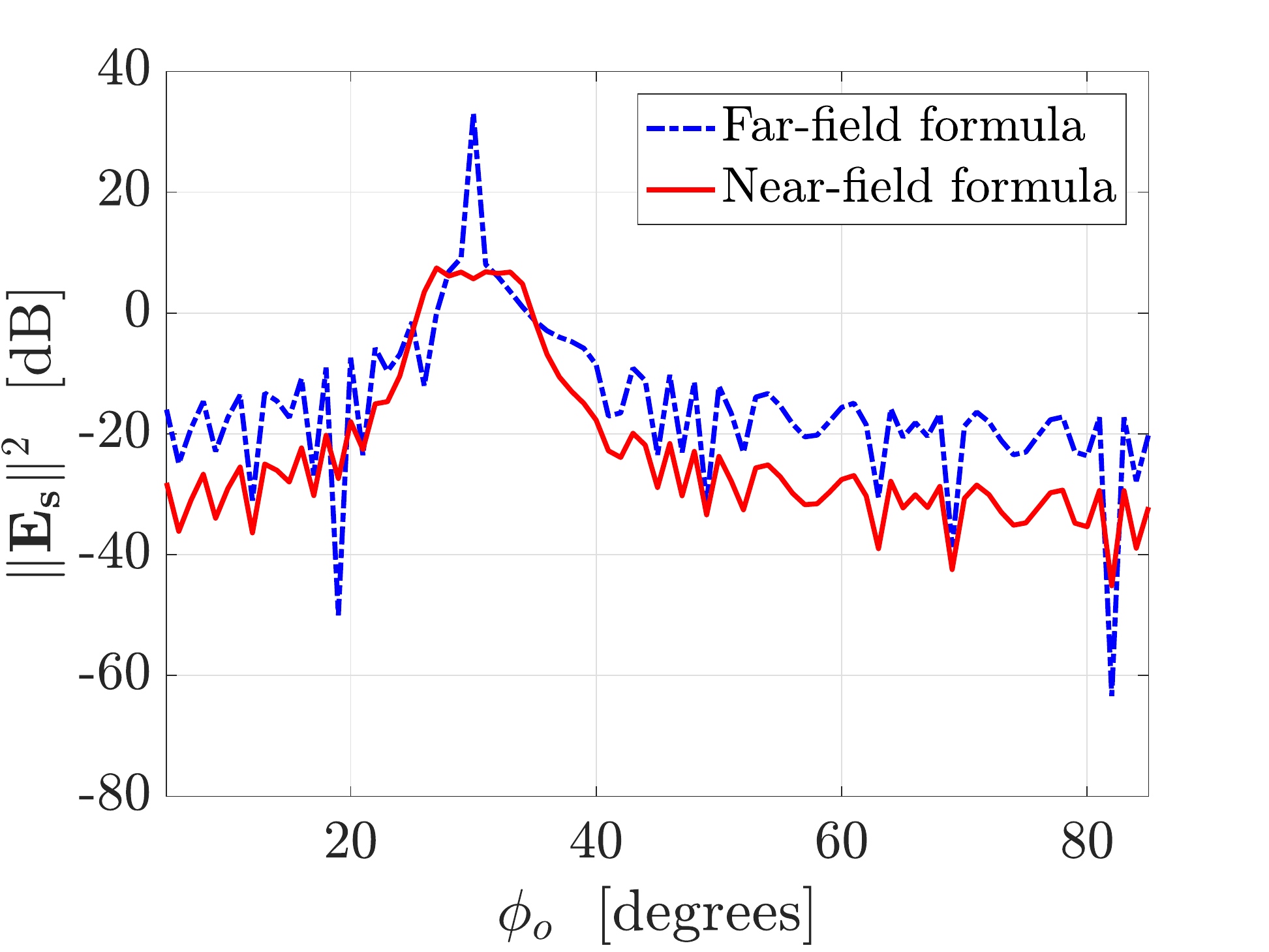}
		\caption{\footnotesize $r_r=2$ m and $r_o = 8$ m}
		\label{fig:Fig_scatteredField1}
	\end{subfigure}%
	\begin{subfigure}{.5\textwidth}
		\centering
		\includegraphics[width=0.83\linewidth]{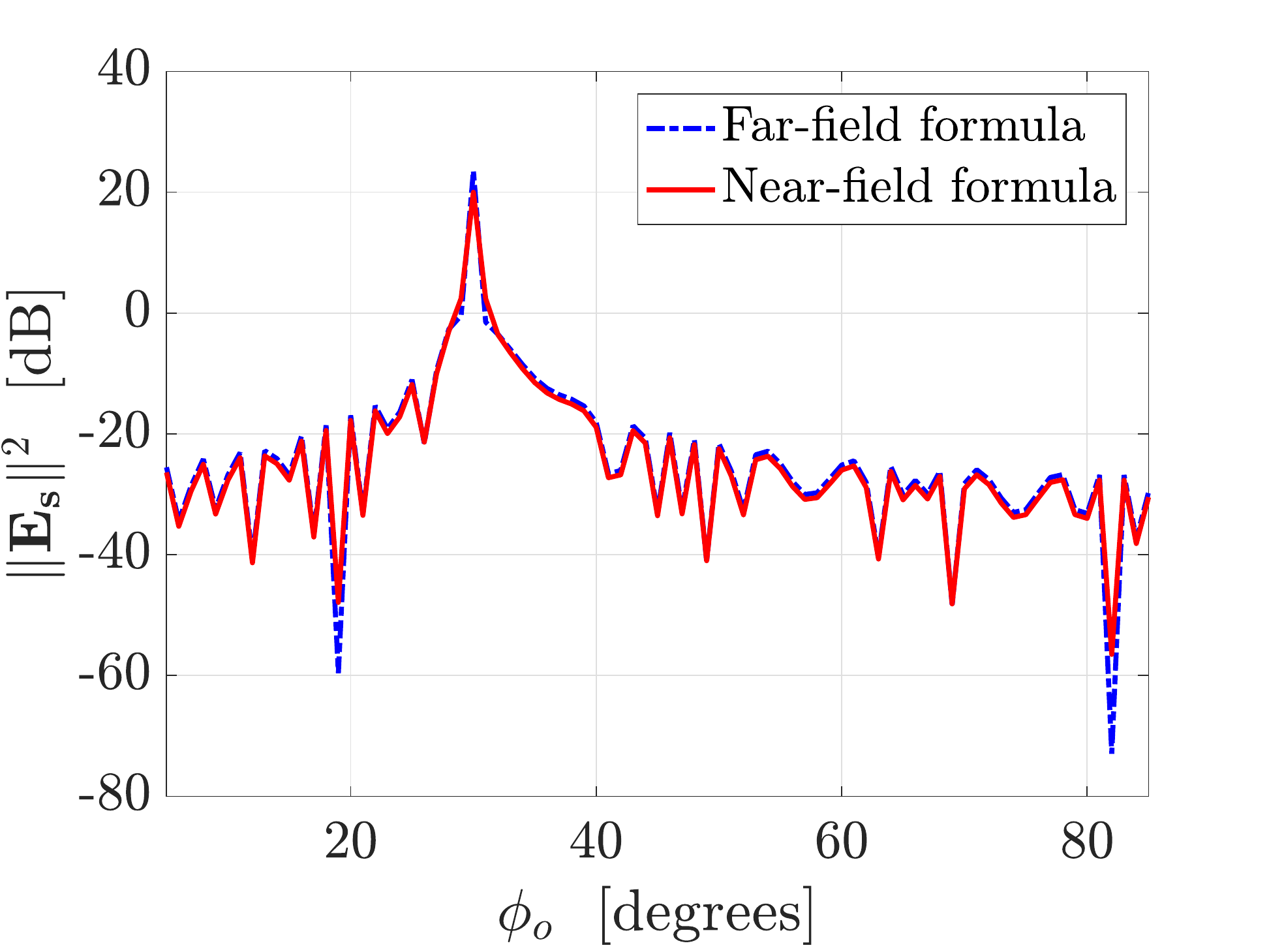}
		\caption{\footnotesize $r_r=6$ m and $r_o = 8$ m}
		\label{fig:Fig_scatteredField2}
	\end{subfigure}%
	\caption{Squared magnitude of the scattered E-field versus observation angle $\phi_o$; $|E_i| = 1$, $L_y=L_z=200\lambda$, $f = 300$~GHz, $\phi_t = 36\degree$, $(\theta_r,\phi_r) = (45\degree,30\degree)$, and $(\theta_o,\phi_o) = (45\degree, \phi_o)$.}
	\label{fig:Fig_scatteredField}
\end{figure*}
\begin{figure*}[b]
	\hrulefill
	\begin{align}\label{eq:space_factor}
	&S_{yz}=\frac{\pi}{4jkL_yL_z\sqrt{a_ya_z}}\left[\text{erf}\left(\sqrt{jka_y}\left(\frac{L_y}{2} - \frac{b_y}{2a_y}\right)\right)- \text{erf}\left(\sqrt{jka_y}\left(-\frac{L_y}{2} - \frac{b_y}{2a_y}\right)\right) \right]\nonumber \\
	&\quad\quad\quad\quad\quad \times  \left[\text{erf}\left(\sqrt{jka_z}\left(\frac{L_z}{2} - \frac{b_z}{2a_z}\right)\right)- \text{erf}\left(\sqrt{jka_z}\left(-\frac{L_z}{2} - \frac{b_z}{2a_z}\right)\right) \right].
	\end{align}
\end{figure*}  
\begin{align}
a_y  &=\frac{(1-\sin^2\phi_t\sin^2\theta_t)}{2r_t} + \frac{(1-\sin^2\phi_r\sin^2\theta_r)}{2r_r}-C_1,  \nonumber\\
b_y & =\sin\phi_t \sin\theta_t + \sin\phi_r \sin\theta_r + C_2, \\
a_z &= \frac{\sin^2\theta_t}{2r_t} + \frac{\sin^2\theta_r}{2r_r}-C_3,\nonumber\\
b_z&=\cos\theta_t + \cos\theta_r +  C_4.
\end{align}
\begin{proof}
See Appendix.
\end{proof}
\end{proposition}
\begin{remark}
In the far-field, the parallel-ray approximations
\begin{align}
\tilde{r}_t(y,z) &\approx - y\sin\phi_t \sin\theta_t - z\cos\theta_t, \\
\tilde{r}_r(y,z) &\approx - y\sin\phi_r \sin\theta_r - z\cos\theta_r
\end{align}
are employed. Then, $a_y = a_z = 0$, and the space factor reduces to~\cite{balanis_electromagnetics}
\begin{align}\label{eq:space_factor_ff}
S_{yz} =\frac{\int_{-L_y/2}^{L_y/2}\int_{-L_z/2}^{L_z/2} e^{j k(b_y y +b_z z)}dydz}{L_yL_z} = \emph{sinc}(Y)\emph{sinc}(Z),
\end{align}
where $Y \triangleq k L_yb_y/2$ and $Z \triangleq kL_zb_z/2$. 	
\end{remark}
From Proposition 1, the squared magnitude of the scattered E-field is calculated as
\begin{align}\label{eq:near_field_formula}
\|\mathbf{E}_{s}\|^2 &= \left(\frac{L_yL_z}{\lambda}\right)^2\frac{|E_i|^2}{r_r^2} \cos^2\phi_t\sin^2\theta_r|S_{yz}|^2,
\end{align}
where $|S_{yz}|^2$ is the normalized beampattern of the \ac{IRS}.

\subsection{End-to-End Signal Model} 
We now introduce the signal model of a holographic \ac{IRS}-assisted \ac{THz} system, where the \ac{Tx} and \ac{Rx} are equipped with a single antenna each. First, recall the relation between the magnitude of the incident wave $|E_i|$ and the transmit power $P_t$, which is $|E_i|^2/\eta = G_tP_t/(4\pi r^2_t)$~\cite{balanis_antenna_theory}. Hence, the power density (W/m$^2$) of the scattered field is 
\begin{equation}
S_s =  \frac{\|\mathbf{E}_s\|^2}{\eta} =  \left(\frac{L_yL_z}{\lambda}\right)^2\frac{P_tG_t}{4\pi r^2_tr_r^2} \cos^2\phi_t\sin^2\theta_r|S_{yz}|^2.
\end{equation} 
Considering the \ac{Rx} antenna aperture $A_r = G_r \lambda^2/(4\pi)$ yields the received power $P_r = S_s A_r$. Lastly, taking into account the molecular absorption loss at \ac{THz} frequencies results in the path loss of the \ac{Tx}-IRS-{Rx} link
\begin{align}\label{eq:pl_expression}
\overline{\text{PL}} &= G_tG_r\left(\frac{L_yL_z}{4\pi}\right)^2\frac{\cos^2\phi_t\sin^2\theta_r}{r_t^2r_r^2}e^{-\kappa_{\text{abs}}(f)(r_t + r_r)}|S_{yz}|^2 \nonumber \\
& = \text{PL}|S_{yz}|^2,
\end{align}
where $\kappa_{\text{abs}}(f)$ denotes the molecular absorption coefficient at the carrier frequency $f$. From~\eqref{eq:pl_expression}, it is evident that the path loss of an \ac{IRS}-assisted link follows the plate scattering paradigm. Combining~\eqref{eq:scattered_efield} and~\eqref{eq:pl_expression}, the baseband signal at the \ac{Rx} is written as
\begin{equation}
y = \left(\sqrt{\text{PL}}e^{-jk(r_r+r_t)}S_{yz} + \sqrt{\text{PL}_d}e^{-jkr_d}\right)s + \tilde{n},
\end{equation}
where $s\sim\mathcal{CN}(0,P_t)$ is the transmitted data symbol, $P_t$ is the average power per data symbol, $r_d$ is the distance between the \ac{Tx} and \ac{Rx}, $\text{PL}_d = G_tG_r\lambda^2/(4\pi r_d)^2e^{-\kappa_{\text{abs}(f)}r_d}$ is the path loss of the direct \ac{Tx}-\ac{Rx} channel, and $\tilde{n}\sim\mathcal{CN}(0,\sigma^2)$ is the additive~noise.

\section{Discussion}
In this section, we discuss in detail the near-field channel model introduced in Section~\ref{sec:channel_model}.
\subsection{Near-Field versus Far-Field Response}
Consider the phase profile \eqref{eq:phase_profile} with
\begin{align}
C_1 &= \frac{1-\sin^2\phi_t\sin^2\theta_t}{2r_t} + \frac{1-\sin^2\phi_o\sin^2\theta_o}{2r_o},\\
C_2 & = - \sin\phi_t \sin\theta_t - \sin\phi_o \sin\theta_o,\\
C_3 &=  \frac{\sin^2\theta_t}{2r_t} + \frac{\sin^2\theta_o}{2r_o},\\
C_4 & =  -\cos\theta_t - \cos\theta_o,
\end{align}
where $(r_o\cos\phi_o\sin\theta_o,r_o\sin\phi_o\sin\theta_o,r_o\cos\theta_o)$ is an arbitrary observation position, with $r_o$, $\phi_o$, and $\theta_o$ denoting the corresponding radial distance, azimuth angle, and polar angle, respectively. Then, the parameters of the beampattern $|S_{yz}|^2$~are
\begin{align}
a_y &= \frac{1-\sin^2\phi_r\sin^2\theta_r}{2r_r} - \frac{1-\sin^2\phi_o\sin^2\theta_o}{2r_o},\\
b_y & = \sin\phi_r \sin\theta_r - \sin\phi_o \sin\theta_o,\\
a_z &= \frac{\sin^2\theta_r}{2r_r} - \frac{\sin^2\theta_o}{2r_o},\\
b_z & = \cos\theta_r -  \cos\theta_o.
\end{align}
We now plot the squared magnitude of the scattered E-field for the considered $\varphi(y,z)$. From Fig.~\ref{fig:Fig_scatteredField}, we first observe that the peak value is at $\phi_o = \phi_r = 30\degree$, as expected. From Fig.~\ref{fig:Fig_scatteredField}(\subref{fig:Fig_scatteredField1}), however, we see a mismatch between the near and far scattered fields of a large IRS. This discrepancy is due to the spherical wavefront of the incident wave, which makes the beampattern $|S_{xy}|^2$ depend on the angles of arrival/departure as well as the distances between the \ac{IRS}, the \ac{Rx}, and the observation point. This unique feature manifests only in the near-field~\cite{irs_em_perspective}. It is finally worth stressing that the near-field space factor in~\eqref{eq:space_factor} coincides with its far-field counterpart \eqref{eq:space_factor_ff} for either an electrically small \ac{IRS} or relatively large distances $r_r$ and~$r_o$, i.e.,~Fig.~\ref{fig:Fig_scatteredField}(\subref{fig:Fig_scatteredField2}).

\subsection{Discrete IRS}
It might be difficult to implement a holographic \ac{IRS} in practice. Therefore, a contiguous \ac{IRS} of size $L_y\times L_z$ can be approximated by a planar array of $N_y = L_y/\tilde{L}_y$ and $N_z = L_z/\tilde{L}_z$ reflecting elements, each of size $\tilde{L}_y\times \tilde{L}_z$; the inter-element spacing is negligible, and hence is ignored. Then,~\eqref{eq:near_field_formula} is recast as
\begin{align}
\|\mathbf{E}_{s}\|^2 = N^2_yN^2_z\left(\frac{\tilde{L}_y\tilde{L}_z}{\lambda}\right)^2\frac{|E_i|^2}{r_o^2} \cos^2\phi_t\sin^2\theta_r|S_{yz}|^2,
\end{align}
where 
\begin{align}\label{discrete_space_factor}
S_{yz} &= \frac{\sum_{n=-\frac{N_y}{2}}^{\frac{N_y}{2}-1} e^{-jk\left((n\tilde{L}_y)^2a_y - n\tilde{L}_yb_y\right)}}{N_y} \nonumber\\
&\times \frac{
\sum_{m=-\frac{N_z}{2}}^{\frac{N_z}{2}-1} e^{-jk\left((m\tilde{L}_z)^2a_z - m\tilde{L}_zb_z\right)}}{N_z},
\end{align}
which follows from~\eqref{eq:equation_beampattern} in the appendix for $y= n\tilde{L}_y$, $z=m\tilde{L}_z$, $L_y=N_y\tilde{L}_y$, $L_z=N_z\tilde{L}_z$, $dy=\tilde{L}_y$, and $dz=\tilde{L}_z$. Likewise, the reflection coefficient of the $(n,m)$th \ac{IRS} element is defined as $e^{j\varphi_{n,m}}$, where $\varphi_{n,m} \triangleq \varphi(n\tilde{L}_y,m\tilde{L}_z)$. For a discrete \ac{IRS}, when the observation direction coincides with that of the \ac{Rx}, $a_y = b_y=a_z=b_z=0$, $S_{yz} = 1$, and a power gain of $(N_yN_z)^2$ is attained over the \ac{Tx}-\ac{IRS}-\ac{Rx} link.
\begin{figure*}[t]
	\centering
	\begin{subfigure}{.5\textwidth}
		\centering
		\includegraphics[width=1.05\linewidth]{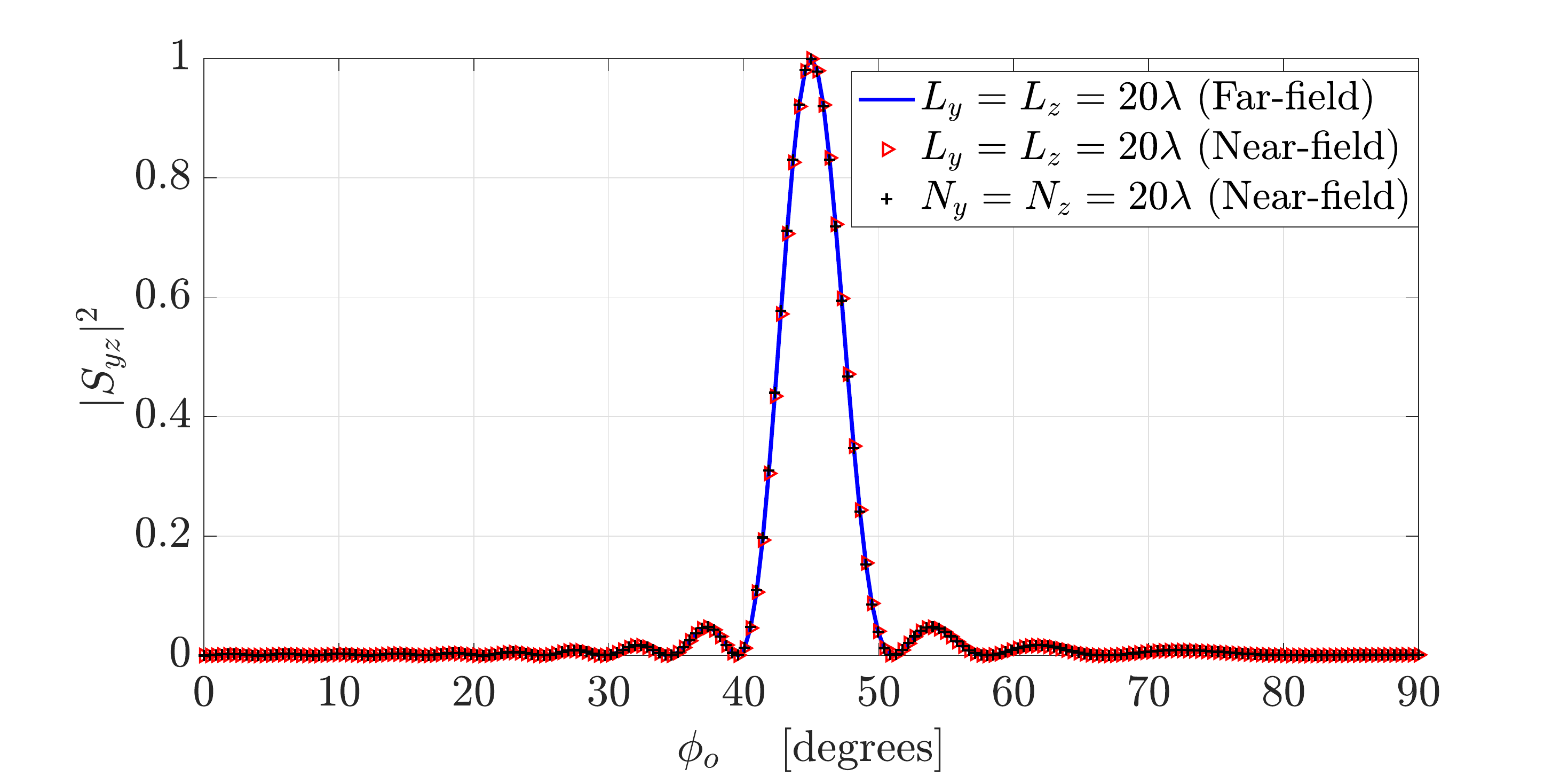}
		\caption{\footnotesize Small IRS}
		\label{fig:Fig_4a}
	\end{subfigure}%
	\begin{subfigure}{.5\textwidth}
		\centering
		\includegraphics[width=1.01\linewidth]{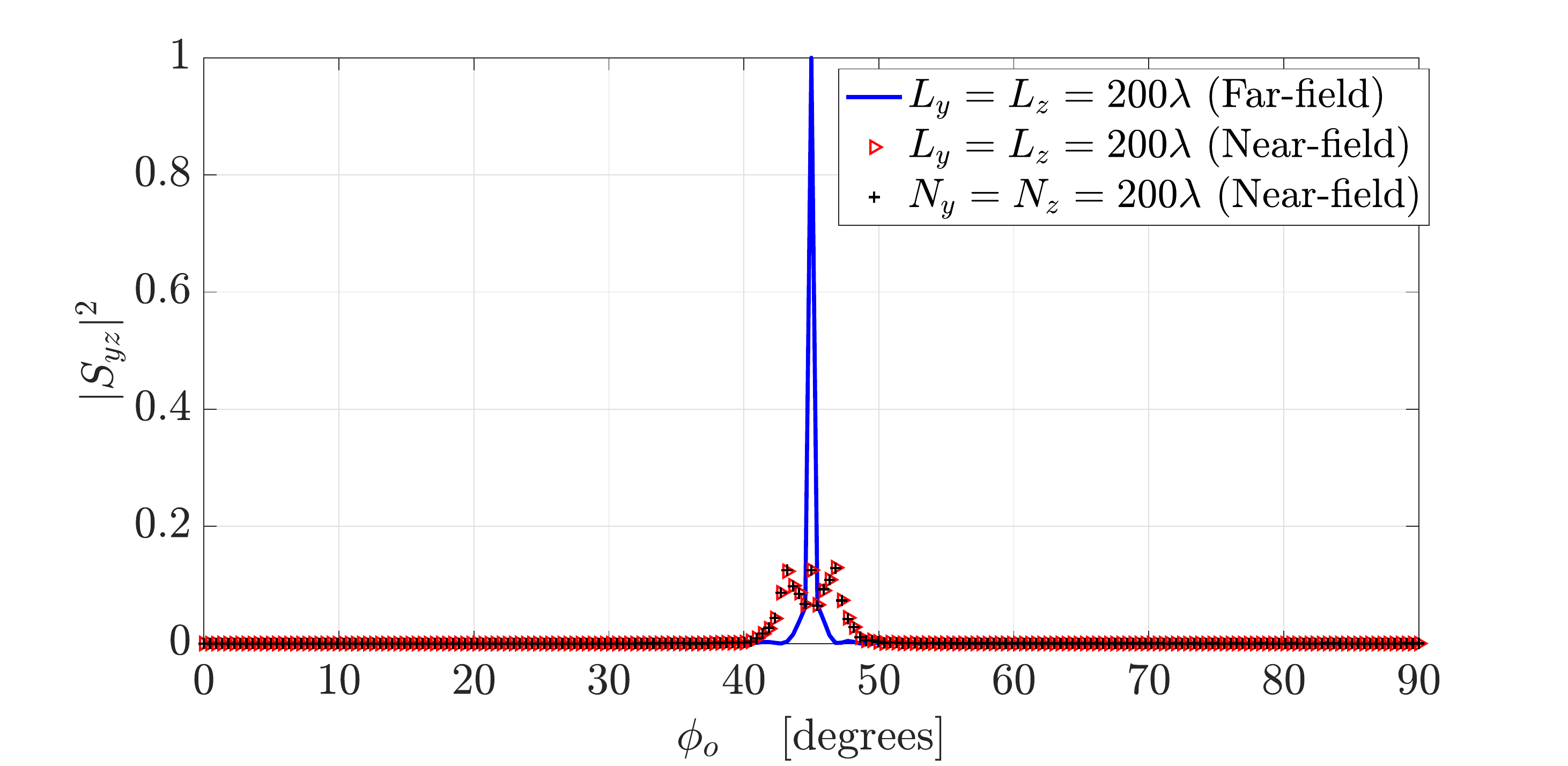}
		\caption{\footnotesize Large IRS}
		\label{fig:Fig_4b}
	\end{subfigure}%
	\caption{Normalized beampattern of holographic \ac{IRS} and discrete \ac{IRS} versus observation angle $\phi_o$; $\tilde{L}_y = \tilde{L}_z = \lambda$, $(r_r,\theta_r,\phi_r) = (2, 45\degree, 45\degree)$, $(r_o,\theta_o,\phi_o) = (8, \theta_o, 45\degree)$, and $ f = 300$ GHz.}
	\label{fig:Fig_beampattern}
\end{figure*}

\subsection{Beamfocusing Capabilities}
With proper design of the phase profile $\varphi(y,z)$, we can cancel out the incident phase and focus the beam into the \ac{Rx} point $(r_r,\theta_r,\phi_r)$.\footnote{This is in sharp contrast to traditional beamforming, where the \ac{IRS} acts as an anomalous reflector that focuses the signal into a desired direction $(\theta_r,\phi_r)$, rather than into a point $(r_r,\theta_r,\phi_r)$~\cite{ris_myths}.} As previously shown, the peak value of $|S_{yz}|^2$ occurs at $(r_o,\theta_o,\phi_o) = (r_r,\theta_r,\phi_r)$. From~Fig.~\ref{fig:Fig_beampattern}(\subref{fig:Fig_4a}) and Fig.~\ref{fig:Fig_beampattern}(\subref{fig:Fig_4b}), we first observe the excellent match between a holographic \ac{IRS} and its discrete counterpart with a negligible inter-element spacing. This implies that we can properly discretize the holographic \ac{IRS} without sacrificing its extremely high spatial resolution. Consequently, \eqref{eq:space_factor} and~\eqref{discrete_space_factor} can be used interchangeably. We further see that the electrically large \ac{IRS} can discriminate two points with the same angular direction $(\theta_o,\phi_o) = (\theta_r,\phi_r)$ but with different distances $r_o\neq r_r$; asymptotically, we have $|S_{yz}|^2 \to 0$ as  $L_yL_z\to\infty$. The beamfocusing capability can be exploited in multi-user transmissions to suppress interference with an unprecedented way. For example, consider an uplink scenario where two users, user~$1$ and user~$2$, simultaneously transmit. Their positions from the \ac{IRS}  are $(r_1,\theta_1,\phi_1)$ and $(r_2,\theta_2,\phi_2)$, with $(\theta_1,\phi_1)=(\theta_2,\phi_2)$ and $r_1\neq r_2$. In the far-field, $|S_{yz}|^2=1$, and hence we will have strong inter-user interference at the \ac{Rx}. Conversely, in the near-field, $|S_{yz}|^2 < 1$ and the inter-user interference becomes small at the \ac{Rx}. 

\subsection{Scattering versus Antenna-Based Path Loss Models}
\begin{figure}[t]
	\centering
	\includegraphics[width=0.82\linewidth]{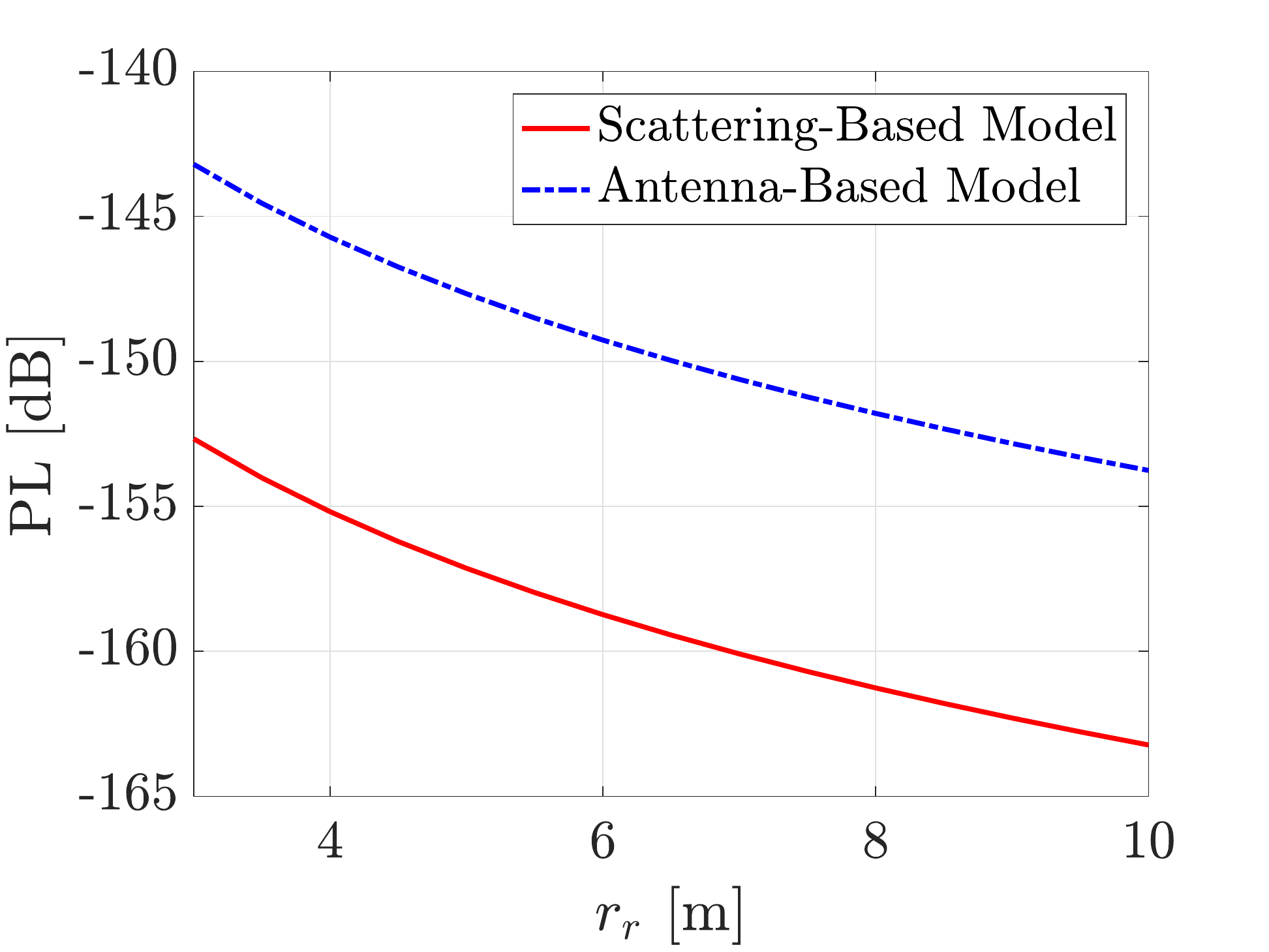}
	\caption{In the antenna-based model~\cite{nf_pathloss_model1}, $G_e(\theta) = \gamma \cos^{2q}\theta$, with $\gamma = \pi$ and $q = 0.285$. The other parameters are: $f = 300$~GHz, $L_y = L_z = \lambda/2$, $r_t=2$ m, $(\theta_t,\phi_t) = (60\degree, 90\degree)$, $(\theta_r,\phi_r) = (45\degree, 90\degree)$, $G_t = 20$~dBi, $G_r = 0$ dBi, and $\kappa_{\text{abs}}(f) = 0.0033$~m$^{-1}$.}
	\label{fig:Fig_PL_antennatheory}
\end{figure}

Some works in the literature (e.g.,~\cite{nf_pathloss_model1}) treat an \ac{IRS} element as a standard antenna that re-radiates the impinging wave. In this case, the path loss is calculated as
\begin{equation}
\text{PL}' = G_tG_r\left(\frac{\lambda}{4\pi}\right)^4\frac{G_e(\theta_t)G_e(\theta_r)}{r^2_tr^2_r}e^{-\kappa_{\text{abs}}(f)(r_t + r_r)},
\end{equation}	
where $G_e(\cdot)$ is the radiation pattern of each \ac{IRS} element. For a sub-wavelength IRS element, it holds that $|S_{yz}|^2 \approx 1$, and $\text{PL} = G_tG_r\left(\frac{L_yL_z}{4\pi}\right)^2\frac{\cos^2\phi_t\sin^2\theta_r}{r_t^2r_r^2}e^{-\kappa_{\text{abs}}(f)(r_t + r_r)}\neq \text{PL}'$, as shown in Fig.~\ref{fig:Fig_PL_antennatheory}. Consequently, simplistic path loss models may not always capture the unique features of \ac{IRS}-aided propagation.
\section{Conclusions}
We have studied, for the first time, the near-field response of holographic \ac{IRSs} operating at the \ac{THz} frequency band. To have a physics-consistent channel model, we leveraged \ac{EM} theory and derived a novel closed-form expression for the scattered field. Unlike existing works, our model accounts for arbitrary incident and reflection angles. Capitalizing on our analysis, we then compared the near-field response with its far-field counterpart and revealed a significant discrepancy, which makes the use of the former necessary for electrically large \ac{IRS}s. We finally discussed the beamfocusing property, which manifests on the near-field regime, and highlighted its potential in multi-user transmissions and interference suppression. For future work, it would be interesting to study the coupling effects in ultra-dense discrete \ac{IRS}s and their connection with super-directive antenna arrays. Moreover, it would be interesting to derive a circuit theory-based model for the power consumption of \ac{THz} \ac{IRSs}. 

\section*{Acknowledgements}
This project has received funding from the European Research Council (ERC) under the European Union’s Horizon 2020 research and innovation programme (grant agreement No. 101001331).
\section*{Appendix}
The magnetic field in~\eqref{eq:H_field_spherical} is written in Cartesian coordinates~as
\begin{equation}
\mathbf{H}_i = -H_{\phi}\sin\phi_t \mathbf{e}_x + H_{\phi}\cos\phi_t \mathbf{e}_y.
\end{equation}
The current density induced on the \ac{IRS} is
\begin{align}\label{eq:surface_current_density}
&\mathbf{J}(y,z) = (2\mathbf{e}_x\times \mathbf{H}_i|_{x=0}) e^{j\varphi(y,z)}\nonumber\\ 
&= 2H_{\phi}\cos\phi_t e^{j\varphi(y,z)} \mathbf{e}_z \nonumber\\ 
& = \frac{2j}{\eta}\sqrt{\frac{\eta P_tG_t}{4\pi}}\frac{e^{-jkr_t}}{r_t}e^{-jk\tilde{r}_t(y,z)}\cos\phi_t e^{j\varphi(y,z)} \mathbf{e}_z  \nonumber\\ 
&=\frac{2j}{\eta}E_ie^{-jk\tilde{r}_t(y,z)}\cos\phi_t e^{j\varphi(y,z)}\mathbf{e}_z\nonumber\\ 
& = J_ze^{j\varphi(y,z)} \mathbf{e}_z,
\end{align}
where $E_i = \sqrt{\frac{\eta P_tG_t}{4\pi}}\frac{e^{-jkr_t}}{r_t}$. Then,~\eqref{eq:Atheta} and~\eqref{eq:Aphi} give
\begin{align}
&\tilde{A}_\theta =- j L_yL_z\frac{2E_i}{\eta}\cos\phi_t\sin\theta_r S_{yz}, \\
&\tilde{A}_\phi = 0,
\end{align} 
where
\begin{align}\label{eq:equation_beampattern}
S_{yz} &= \frac{\int\!\!\int_\mathcal{S} e^{-j k(\tilde{r}_t(y,z) + \tilde{r}_r(y,z) - \varphi(y,z)/k)}ds}{L_yL_z} \nonumber \\
&=\frac{\int_{-L_y/2}^{L_y/2}\int_{-L_z/2}^{L_z/2} e^{-j k(a_yy^2 -b_yy +a_zz^2-b_zz)} dydz}{L_yL_z},
\end{align}
with 
\begin{align}
a_y  &=\frac{(1-\sin^2\phi_t\sin^2\theta_t)}{2r_t} + \frac{(1-\sin^2\phi_r\sin^2\theta_r)}{2r_r}-C_1,  \nonumber\\
b_y & =\sin\phi_t \sin\theta_t + \sin\phi_r \sin\theta_r + C_2, \\
a_z &= \frac{\sin^2\theta_t}{2r_t} + \frac{\sin^2\theta_r}{2r_r}-C_3,\\
b_z&=\cos\theta_t + \cos\theta_r +  C_4.
\end{align}
We now use the identity
\begin{equation}\label{eq:erf_identity}
\int e^{-jk(ay^2-by)} dy = \frac{\sqrt{\pi}}{2 \sqrt{jka}}\text{erf}\left(\sqrt{jka}\left(y - \frac{b}{2a}\right)\right), 
\end{equation}
which follows from the definition of the error function, some algebrain manipulations, and a change of variables. Using~\eqref{eq:erf_identity}, the expression~\eqref{eq:space_factor} for $S_{yz}$ is derived. The scattered E-field is finally given by 
\begin{align}
&\mathbf{E}_s  = -\eta \frac{j k e^{-kj r_r}}{4\pi r_r} (\tilde{A}_\theta\mathbf{e}_\theta + \tilde{A}_\phi \mathbf{e}_\phi) \nonumber\\
&=\frac{j^2 L_yL_z k E_i e^{-kj (r_t + r_r)}}{2\pi r_r} \cos\phi_t\sin\theta_r S_{yz}\mathbf{e}_{\theta}  \nonumber\\
& = -\frac{L_yL_z}{\lambda}\frac{|E_i| e^{-kj (r_t + r_r)}}{r_r}\cos\phi_t\sin\theta_r S_{yz} \mathbf{e}_{\theta},
\end{align}
which completes the proof.

\end{document}